# Pressure tuning the lattice and optical response of silver sulfide

Zhao Zhao[1*], Hua Wei[2], and Wendy L. Mao[3,4]

[1] *Department of Physics, Stanford University, Stanford, CA 94305, USA*

[2] *Scintillation Materials Research Center, Department of Materials Science and Engineering, University of Tennessee, Knoxville, Tennessee 37996, USA*

[3] *Photon Science and Stanford Institute for Materials and Energy Sciences, SLAC National Accelerator Laboratory, Menlo Park, CA 94025, USA;*

[4] *Department of Geological Sciences, Stanford University, Stanford, CA 94305, USA*

Electronic address: zhaozhao@stanford.edu

Binary transition metal chalcogenides have attracted increasing attention for their unique structural and electronic properties. High pressure is powerful tool for tuning their crystal structure and electronic structure away from their pristine states. In this work, we systematically studied the *in situ* structural and optical behavior of silver sulfide ($Ag_2S$) under pressure by X-ray diffraction (XRD) and Infrared (IR) spectroscopy measurements in a diamond anvil cell. Upon compression, $Ag_2S$ undergoes structural symmetrization from monoclinic to orthorhombic, represented by the decrease of $\beta$ from $99^o$ to $90^o$ through a series of structural transitions coupled with lattice contractions. IR transmission and reflectivity measurements showed that pressure effectively tunes semiconducting $Ag_2S$ into a metal at ~ 22 GPa. Drude model analysis of the IR reflectivity indicates that the optical conductivity evolves significantly, reaching the highest conductivity of 100 $\Omega^{-1}cm^{-1}$ at ~ 40 GPa. Our results highlight pressure's dramatic role in tuning the structural and electronic state of silver chalcogenides and suggest $Ag_2S$ as a platform for developing optical and opto-electronics applications.

Binary transition metal chalcogenides have attracted great research and industrial interests in recent years. For example, the discovery of topological insulators $Bi_2Te_3$ and $Bi_2Se_3$ has opened up novel states of matter for condensed matter research.[1–3] The resurgent interest on layered transition-metal chalcogenides $WX_2$ and $MoX_2$ (X= S, Se, and Te) presents promising opportunities in developing next-generation electronics and transistors.[4–7] Silver chalcogenides $Ag_2X$ is another group of interest. $Ag_2Se$ and $Ag_2Te$ with bulk band-gaps of ~ 0.1 to 0.2 eV were proposed as new 3D topological insulators with a highly anisotropic Dirac cone[8] that may lead to new spintronics applications.[9] Experimental evidence supporting metallic surface states has been found in the case of $Ag_2Te$.[10,11] In contrast, $Ag_2S$ is a semiconductor with band-gap ~ 1 eV.[12,13] $Ag_2S$ devices enable fabrication of quantized conductance atomic switches that can largely overcome the limitation of current semiconductor devices[14] and $Ag_2S$ quantum dots can be applied in high sensitivity cancer imaging.[15]

The ability to tune semiconductors' electronic structures away from their pristine states is fundamental to semiconductor research, because this capability could lead to novel electronics and opto-electronics applications. For $Ag_2X$, a variety of experimental techniques have been employed to engineer their electronic structures and band-gaps. Two prominent examples involve tuning their dimensionality via film thickness and particle size. By varying the thickness[16–20], its activation gap can be effectively tuned.[16] By changing the particle size, its optical band-gap could also be engineered.[17–19]

Compared with the above methods, pressure is a powerful tool in modifying the way atoms arrange and thus to tune materials electronic structure and physical properties. Previous high pressure studies on transition metal chalcogenides have discovered a variety of exciting crystal structures and electronic structures. Studies on $Bi_2Te_3$ and $Bi_2Se_3$ showed pressure induced new crystal structures with different electronic structures. Studies on $MoS_2$ and $MoSe_2$ show that pressure can continuously tune their band-gap and lead to metallization. For $Ag_2X$, earlier work on $Ag_2Te$ and $Ag_2Se$ showed that pressure induced abrupt semiconductor to metal transitions associated with structural transitions at ~ 3 GPa and ~ 8 GPa respectively.[21,22]

For $Ag_2S$, previous studies up to ~ 20 GPa showed it undergoes phase transitions from the monoclinic phase I to an orthorhombic phase II at 5.4 GPa and then to another monoclinic phase III at 10.6 GPa.[23] Interestingly, one high symmetry orthorhombic phase (space group *Pnma*) was predicted at higher pressure. Yet this phase was not observed due to the pressure range in their experiments.[23] Electrical transport measurements up to ~ 19 GPa combined with first-principle calculations showed that $Ag_2S$ remained semiconducting electronically while exhibiting dramatic changes in majority carrier concentration and mobility.[24]

In this work, we aim at solving following tasks: 1) explore possible structural transition at higher pressure, 2) clarify the structural connections among the different phases, 3) and determine how pressure tunes the optical response which relates to the band-gap's and optical conductivity's evolution. We performed *in situ* synchrotron angle dispersive powder X-ray diffraction (XRD) experiments up to ~ 50 GPa and *in situ* synchrotron infrared (IR) spectroscopy measurements up to ~ 40 GPa, see Fig. 1 for schematic set up. Detailed structural and optical analyses are presented.

The $Ag_2S$ sample (purity higher than 99%, product number 241474, Lot number MKBQ6420V) was purchased from Sigma-Aldrich. The sample was checked by XRD, Inductively Coupled Plasma Mass Spectrometry, and X-ray Photoemission Spectroscopy. Symmetric DACs with 300 microns culet size were used. The basic set up for high pressure measurements is shown in Fig. 1. Tungsten thin foil was used as the gasket and a 120 um diameter sample chamber was drilled in the center. The gasket was pre-indented to a thickness of about 40 μm. Ruby spheres were used for determining pressure in all measurements. For XRD measurements, $Ag_2S$ was ground into powder and loaded into the sample chamber. Silicone oil was used as the pressure transmitting medium for XRD measurements.

Two sets of *In situ* angle dispersive XRD data were collected at 12.2.2 of Advanced Light Source (ALS), Lawrence Berkeley National Laboratory (LBNL) with wavelengths of 0.3775 Å and 0.4959 Å. The diffraction images were collected by MAR 345 detector. The 2D diffraction images were integrated by using the program fit2D.[25] Jade is used for space group indexing.[26] Rietveld fitting on integrated 1D powder diffraction pattern was performed by GSAS and EXPGUI package.[27]

The Birth-Murnaghan (BM) Equation of State (EOS) is employed to fit experimental P-T relation, using software EOSFit V5.2.[28] For third-order BM EOS, following equation applies: $P = 3 K_0 f_E (1 + 2f_E)^{5/2} \left( 1 + \frac{3}{2} (K' - 4) f_E \right)$   $f_v = [(V_0/V)^{2/3} - 1]/2$   $V_0$. $K_0$, and $K'$ are fitted parameters. For second-order BM EOS, the $K'$ is fixed at 4. The uncertainties of volumes given by GSAS are included in the fitting for all four phases.

For IR measurements, KBr was used as the pressure transmitting as it is IR transparent. The $Ag_2S$ sample was first squeezed into a 30 μm × 30 μm × 3 μm (thickness) thin foil, then sandwiched between the pressure transmitting medium (KBr) and one side of the culet. Type Ia diamonds generally exhibit strong absorption at 1100-1400 $cm^{-1}$ (0.14 eV-0.17 eV) range and below because of absorption from N impurities. To avoid such absorption, type IIa diamonds almost devoid of impurities were used for IR measurements.[29] High-pressure IR measurements were conducted in beamline U2A of the National Synchrotron Light Source (NSLS), Brookhaven National Laboratory (BNL). Infrared microspectroscopy was performed on a Bruker Vertex 80v FT-IR spectrometer coupled to a Hyperion-2000 microscope with a MCT mid-band detector. Both reflectance and transmittance were measured in the spectral range between 600 and 8000 $cm^{-1}$ with the resolution of 4 $cm^{-1}$.

The high brightness and controllable focus at micro scale allow accurate data calibration and normalization.[30] To obtain transmittance $T(\omega)$ of $Ag_2S$ at each pressure, one transmission measurement was performed at $Ag_2S$ thin foil $I_T^S(\omega)$ and the other at KBr medium nearby $I_T^M(\omega)$. The normalized transmittance $T(\omega) = I_T^S(\omega)/ I_T^M(\omega)$. For IR reflectivity measurement on a DAC, the reflectivity is measured at the $Ag_2S$/diamond interface instead of the $Ag_2S$/vacuum interface. There are four quantities that need to be measured. During high pressure measurement with sample, intensity reflected over sample and diamond $I_R^S(\omega)$ and intensity reflected by the external face of the diamond $I_R^{D1}(\omega)$ were taken. At end of measurement, a gold foil (mirror) was placed to replace the sample where we measured reflectivity between diamond anvil $I_R^{Au}(\omega)$ and by the external face of diamond $I_R^{D2}(\omega)$. $R(\omega) = [I_R^S(\omega) \times I_R^{D1}(\omega)] / [I_R^{Au}(\omega) \times I_R^{D2}(\omega)]$.

To convert reflectivity from $Ag_2S$/diamond interface to $Ag_2S$/vacuum interface, the correction according to Fresnel's formula is employed.[31] $R(\omega)$ at the interface $R(\omega) = [(n - n_0)^2 + k^2]/[(n - n_0)^2 + k^2]$, where $n$ and $k$ are the real and imaginary parts of the complex refractive index of $Ag_2S$ and $n_0 = 2.4$ is the refractive index of diamond. To calculate the complex refractive index of $Ag_2S$, a Drude-Lorentz model was applied for IR reflectivity data using the iterative fitting in Software RefFit.[32] The optical conductivity was determined by the obtained refractive index using the Kramers-Kronig relation.

Fig. 2 shows the four crystal structures of $Ag_2S$ and collected XRD pattern during compression. The lowest pressure measurement at 0.8 GPa confirms that ambient structure. When increasing pressure to 5.1 GPa, phase II appear in coexistent with III. At 8.8 GPa, new peaks from phase III appear. At 28.4 GPa, some peaks merge together, suggesting the symmetrization of phase III to phase IV, see the inset of Fig. 2. Decompression cycles show that all structural transitions are reversible. Representative

Rietveld refinement profiles and crystallographical information are show in the Supplementary Materials, Fig. S1 and Table S1.[33]

By combing the results of Rietveld refinements, we obtained the evolution of unit cell volume and cell parameters in Fig. 3. The structure of phase I has a strong distorted anti-PbCl$_2$ type structure (Fig. 2(a)), where Ag1 and Ag2 atoms are both four-fold coordinated. Ag1 is closer to one S layer on (010) plane, while Ag2 is more in between the two S layers[34,35]. We observe the decrease of $\beta$ angle at around 5 GPa as it changes from around 99° to 98°, and the concurring change of compressibility in $c$ direction. The changes can explained by that before 5 GPa of Ag-S distance decreases fast while Ag-Ag distance almost remains constant, while after 5 GPa all atomic distances decrease at similar rates.[23] Fig. 3(b) shows that among three axes, $b$ direction is the most compressible. The large $K'$17 (2) can be explained by change of compressibility of the monoclinic structure.

Phase II is an interesting structure. It is an orthorhombic (space group $P2_12_12_1$) and isostructural to the ambient structure of Ag$_2$Se. [23] Note that pure phase II was not observed in neither run of experiment. This supports previous calculations showing that phase I and phase II are energetically close. And any disturbance such as small deviation stress or non-hydrostaticity may render phase II unfavorable to phase I. In phase II, two Ag sites are still observed within the structure. S layers are formed at (001) plane, and Ag1 is still closer to the plane when compared with Ag2. Ag1 is four-fold coordinated to form strong distorted tetrahedral. Differently, Ag2 increases from four-fold to five-fold coordinated to form pyramid structure.

The $P2_1/n$ phase III structure is isostructural to phase I structure, seen in Fig. 2(a). In terms of cell parameters, the $a$ and $c$ expanded while $b$ contracted from phase I to phase III, The coordination of Ag2 is maintained as five-fold pyramid structure, but the coordination of Ag1 increase from four to five as Ag1 is almost on the (010) plane formed by S to form trigonal bipyramidal. The *c* axis's incompressibility has close relation to the strong decrease of *β* angle. As pressure increases, strong distortions of Ag-S polyhedral occur due to increasing ionic interactions.

When pressure reached ~ 28 GPa, we for the first time experimentally discovered a *Pnma* phase IV. We tested different structural models among binary chalcogenides and the Pnma structure shows the best fit. The symmetrization from phase III to phase IV is supported from the continuous decrease of *β* to ~ 90° and continuous lattice contractions. In this high symmetry phase IV, two different Ag sites persist where Ag1 form trigonal bipyramidal and Ag2 forming pyramid with nearby S atoms. It shall be noted the crystallographical differences of the two sites persist to the highest pressure.

The lattice response of $Ag_2S$ under pressure will inevitably change its electronic structure, and thus its optical properties. We then performed IR reflectivity and transmission spectroscopy measurements up to ~ 40 GPa. Fig. 4 presents the measured synchrotron IR transmittance spectra and analysis. At the lowest pressure 0.1 GPa, the spectrum (Fig. 4a) shows an abrupt change at the absorption edge ~ 0.9 eV and an IR transmission window at energy region below it. As pressure increases, the transmission window keeps collapsing into lower energy region, which indicates the band-gap's narrowing. At 22.4 GPa, nearly zero transmission is observed in between 0.1 to 1.0 eV, suggesting the metallization. Further increasing pressure to the highest pressure 40.1 GPa does not change the IR spectrum. We can also interpret the IR data by the optical density (OD or $A_\lambda$), where $A_\lambda$ is defined as logT (T as the transmittance). As seen from Fig. 4(b) for the pressure-photon-energy-$A_\lambda$ map, a crossover of low $A_\lambda$ (semiconducting state) to high $A_\lambda$ (metallic state) occurs at around 20 GPa.

For more detailed analysis on the electronic structure, we employed empirical models to fit the optical band-gaps. For an direct band-gap semiconductor, the absorption coefficient is proportional to the root of the energy difference of the photon energy and band-gap, $\alpha \propto (hv - E_{id})^{1/2}/hv$. For an indirect band-gap semiconductor, a square law is followed instead, $\alpha \propto (hv - E_d)^2/hv$. We therefore perform linear extrapolations of $(A_\lambda hv)^{1/2}$ and $(A_\lambda hv)^2$ to determine the direct band-gap $E_d$ and indirect band-gap $E_{id}$. Representative extrapolations are shown in Fig. S2. As pressure increases, $E_{id}$, $E_d$, and $E_c$ unanimously decrease, supporting the band-gap narrowing. Consider that Ag$_2$S is calculated to be a direct band-gap semiconductor, the $E_d$ should more accurately represent the electronic band-gap of Ag$_2$S. At pressure above 22 GPa, the $E_d$ becomes zero effectively.

To better probe the optical response as a function of pressure, we simultaneously measured the IR reflectivity at each pressure (Fig. 5(a)). At below 9 GPa, the spectrum appears flat and featureless. At 9.2 GPa, the reflectivity starts to transform where the low energy part rises up. At 22.4 GPa, the Drude mode, a strong evidence of metallization, can be seen. To quantitatively characterize the optical response, we derived the optical conductivity σ from the Kramers-Kronig analysis. For consistency, a Drude model was used in all of the fittings. The results are shown in Fig. 5(b) and a clear crossover from semiconducting to metallic can be seen. We further plot the zero frequency (DC) conductivity of Ag$_2$S from the fitting. Clear features can be seen: the DC conductivity appears at above 9 GPa with the values of 30 to 40 $\Omega^{-1}$cm$^{-1}$. From 18 to 26 GPa, the DC conductivity strongly increase to about 100 $\Omega^{-1}$cm$^{-1}$ and maintains this value up to the highest pressure studied.

An important question that worths discussion is whether the metallization of Ag$_2$S relate to any structural transition. Phase II to III transition can be ruled out because it occurs at far below the metallization pressure. However, phase III to IV transition occurs between 25 to 28 GPa, just slightly higher than the electronic transition at 22 GPa. We propose two possibilities here. One scenario is that the metallization of Ag$_2$S accompanies the phase III to IV transition, where the difference of pressure mediums

(KBr in the IR experiments and silicone oil in the XRD experiments) lead to the apparent difference in the simultaneous phase III to IV and metallization transition. The other scenario is the metallization occurs within phase III. During the structural symmetrization, the band-gap of phase III decreases and reaches zero. Consider that the structural evolution from phase III to IV is smooth (Fig. 3(b)), the electronic structure of phase IV may be also similar to pressurized phase III and thus be metallic. In either case, the metallization of $Ag_2S$ is a sluggish process because of the continuous structural change.

Among the group of silver chalcogenides ($Ag_2X$), pressure tends to induce a serious of structural transitions. For $Ag_2Se$ and $Ag_2Te$, the abrupt semiconductor to metal transition relates to first-order structural transitions with large changes in crystal structures. However, in the case of $Ag_2S$, the band-gap varies smoothly when undergoing structural transitions. This suggests that among $Ag_2X$, $Ag_2S$ may be a more suitable materials systems for developing optical and opto-electronics applications among $Ag_2X$. Another point worth discussion is that the phase II of $Ag_2Se$ is isostructural to ambient structure of $Ag_2S$ and mostly likely to have similar electronic structure. Considering $Ag_2Se$ is a potential topological insulator with inverted band structure and $Ag_2S$'s band gap becomes as small as ~ 0.4 eV at ~ 5 GPa, there lies the possibility that phase II of $Ag_2S$ is also a topological insulator. Future calculations are encouraged to test this novel scenario.

In summary, we comprehensively studied the high pressure structural and optical behavior of $Ag_2S$ by *in situ* angle dispersive XRD and IR experiments. A new orthorhombic *Pnma* phase IV was experimentally identified for the first time. The structural connections and symmetrization from phase I to phase IV are clearly presented. Pressure continuously tuned semiconducting $Ag_2S$ into a metal at ~ 22 GPa. The optical conductivity shows radical evolution and the highest conductivity reaches 100 $\Omega^{-1}cm^{-1}$ at ~ 40 GPa. Our results highlight pressure's dramatic role in tuning the structural and electronic state of silver chalcogenides and suggest further optical applications based on Ag$_2$S.

FIG. 1. (single-column) Experimental set up for high pressure X-ray and IR measurements. For XRD and IR measurements, silicone oil and KBr were used as the pressure medium respectively. In the IR measurements, a Ag$_2$S thin foil was sandwiched by KBr and the upper culet. Both reflectivity and transmission were measured.

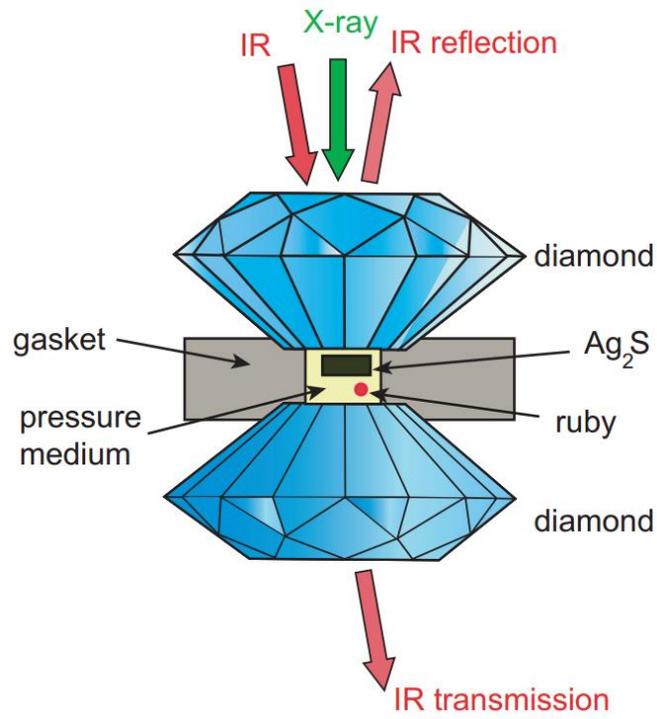

FIG. 2. (single-column) a) Crystal structures for the four phases of $Ag_2S$ under pressure, where two distinctive Ag sites are observed in all structures. b) Representative high pressure XRD patterns with $\lambda = 0.3775$ Å. Numbers on the right are pressures in units of GPa. New diffraction peaks in phase II are shown by stars and phase III by diamonds. The inset shows the symmetrization from phase III to phase IV.

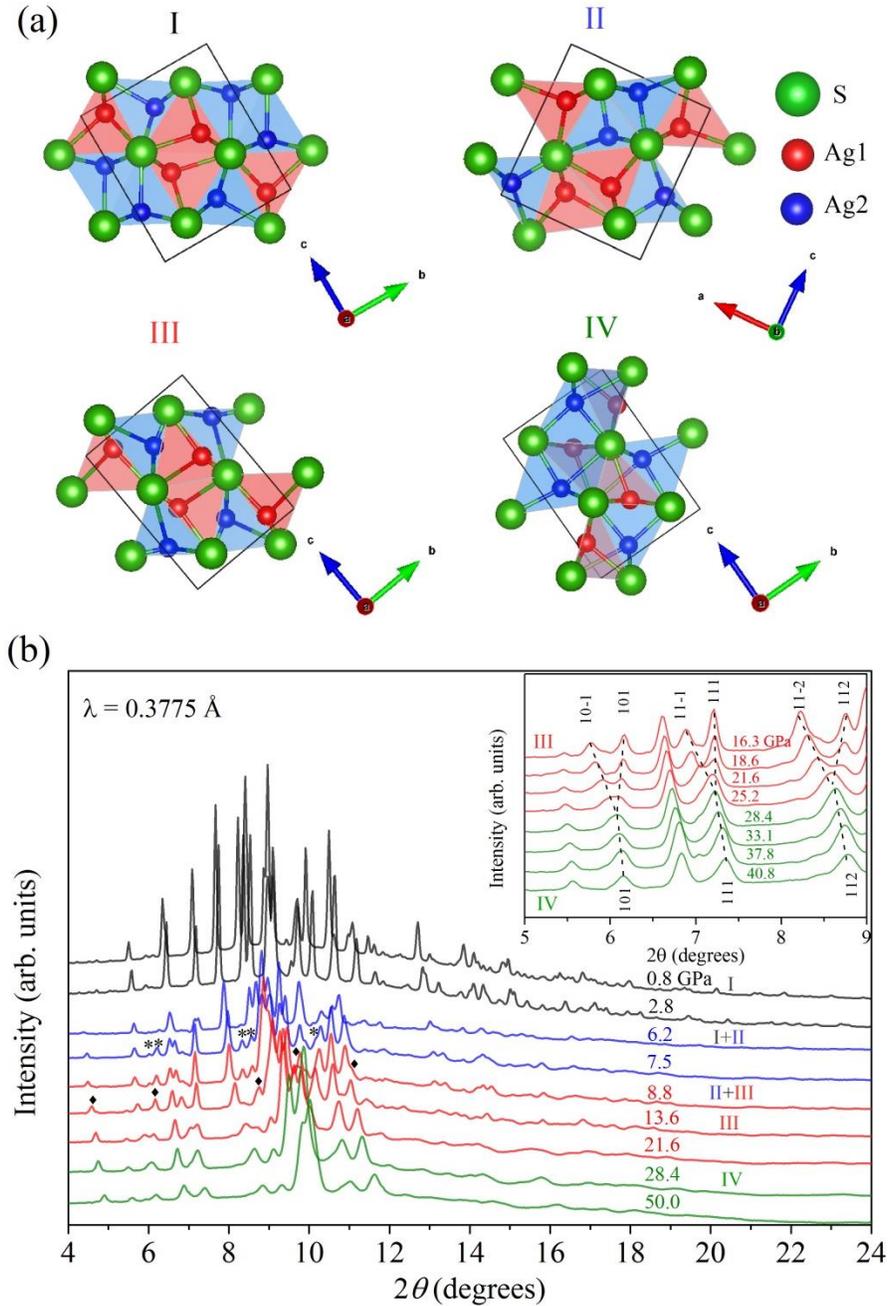

FIG. 3. (single-column) (a) Pressure-volume EoS of $Ag_2S$. Circles and squares represent data from two separate runs. The curves are BM-EOS of each phase where phase I was fit by third-order equation and the rest by second-order equations. (b) Normalized cell parameters and angle for phase I, III, and IV as a function of pressure. Phase II is not plotted here because it has an orthorhombic structure and does not fit in with the continuous decrease of angle. Phase V is converted from *Pnma* to *Pmnb* for better comparison. Uncertainty given by GSAS-EXPGUI is smaller than marker size.

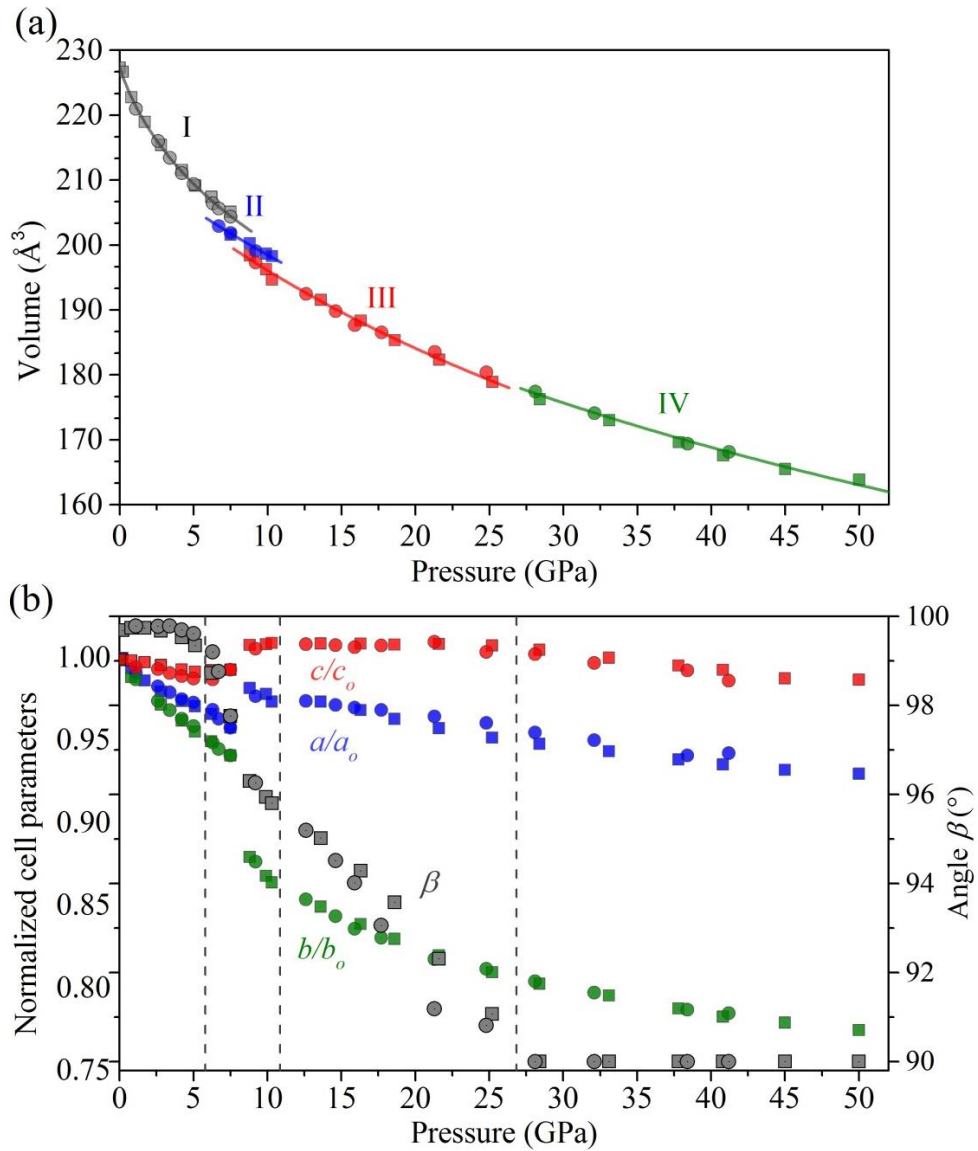

FIG. 4. (double-column) (a) Representative IR transmittance spectra at various pressures. The 0.23-0.28 eV region is obscured by absorption of the diamonds in the DAC. (b) Pressure-photon energy-optical density mapping. (c) Evolution of energy gaps with pressure.

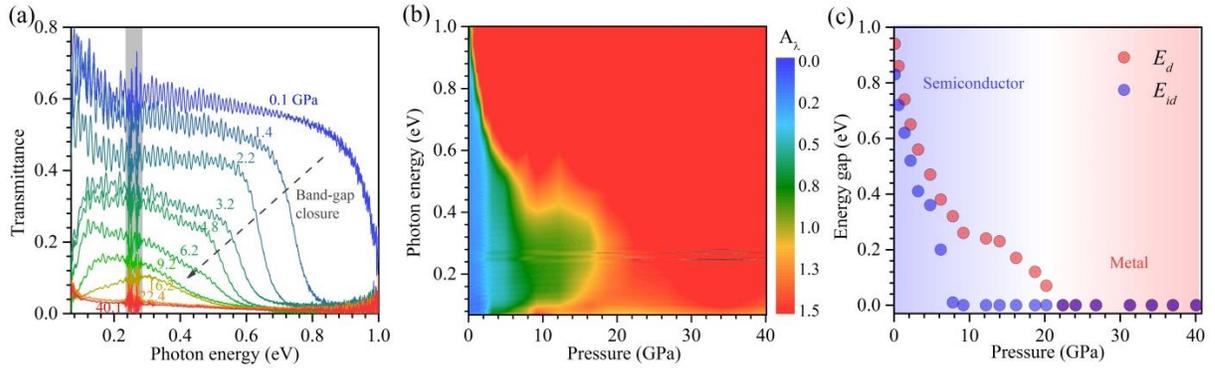

FIG. 5. (double-column) (a) Representative IR reflectance spectra at various pressures. The 0.23-0.28 eV region is obscured by absorption of the diamonds in the DAC. The 0.5-1 eV region is not shown because of its lack of feature. (b) Pressure-photon energy-optical conductivity ($\sigma$) mapping. (c) Evolution of DC conductivity with pressure.

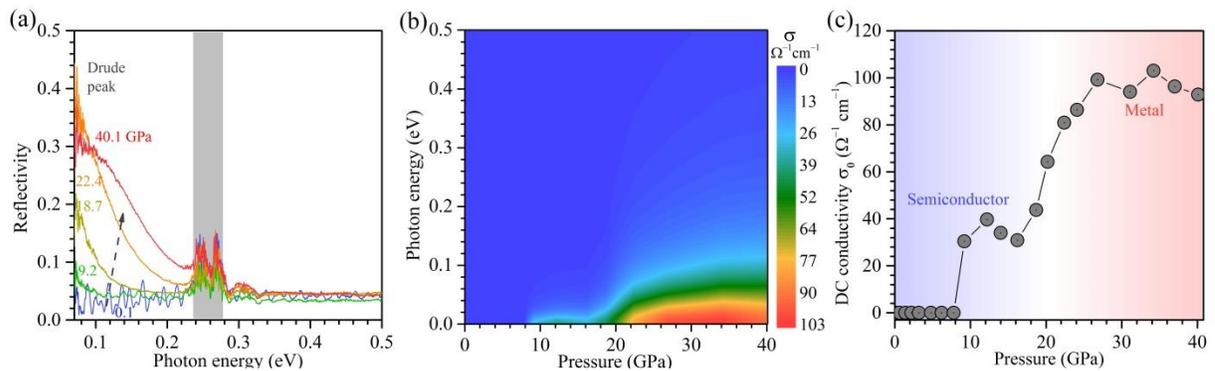